\documentstyle[11pt,newpasp,twoside]{article}
\def\whzsr{\,{\rm W\, Hz^{-1} sr^{-1}}}
\def\edcomment#1{\iffalse\marginpar{\raggedright\sl#1\/}\else\relax\fi}
\reversemarginpar

\begin{document}
\title{Clues on the nature of low-z radio sources from the 2dFGRS}
 \author{M.Magliocchetti + 2dFGRS Team }
\affil{S.I.S.S.A., Via Beirut 2/4, 34100 Trieste, Italy}
\begin{abstract}
We use redshift determinations and spectral analysis of galaxies in the 
2dF Galaxy Redshift Survey to study the properties of local radio sources 
with $S\ge 1$~mJy.
\end{abstract}
\section{The Datasets}
The 2dF Galaxy Redshift Survey (2dFGRS\footnote[1]{{\bf The 2dFGRS Team:}
S.J.\ Maddox (Nottingham), C.A.\ Jackson (ANU), J.\ Bland-Hawthorn
(AAO), T.\ Bridges (AAO), R.\ Cannon (AAO), S.\ Cole (Durham), M.\
Colless (ANU), C.\ Collins (LJMU), W.\ Couch (UNSW), G.\ Dalton
(Oxford), R.\ de Propris (UNSW), S.P.\ Driver (ANU), G.\ Efstathiou
(Cambridge), R.S.\ Ellis (Caltech), C.\ S.\ Frenk (Durham), K.\
Glazebrook (JHU), O.\ Lahav (Cambridge), I.\ Lewis (Oxford), S.\
Lumsden (Leeds), J.A.\ Peacock (Edinburgh), B.A.\ Peterson (ANU), W.\
Sutherland (Edinburgh), K.\ Taylor (Caltech)}; Maddox, 1998) is a 
large-scale survey aimed at obtaining spectra for 250,000 galaxies 
to an extinction-corrected limit for completeness of $b_J=19.45$ over 
an area of 2151 square degrees. 
Redshifts for all the sources brighter than $b_J=19.45$ are determined
in two independent ways, via both cross-correlation of the spectra
with specified absorption-line templates and by
emission-line fitting. The success rate in redshift
acquisition is estimated about 95 per cent.  
The median
redshift of the galaxies is 0.11 and the great majority have $z<0.3$.

Radio observations come from the FIRST (Faint Images of 
the Radio Sky at Twenty centimetres) survey (Becker et al. 1995), 
estimated to be 95 per cent 
complete at 2~mJy and 80 per cent complete at 1~mJy. 
Optical counterparts for a sub-sample of FIRST sources have then been 
obtained by matching together objects included in the radio catalogue with 
objects coming from the APM survey in the overlapping region $9^h 48^m \la 
{\rm RA}({\rm 2000}) \la 14^h 32^m$ and $-2.77^\circ \la {\rm dec}({\rm 2000}) 
\la 2.25^\circ$.  
Out of approximately 24,000 radio sources with $S\ge 1$~mJy in the
considered area, Magliocchetti \& Maddox (2001) find 4075
identifications in the APM catalogue for $b_J\le 22$ (the limiting magnitude 
of the APM survey) and 971 for $b_J\le 19.45$ (the completeness limit of the 
2dFGRS), both obtained for a matching radius of 2 arcsecs.
2dF data then provided optical spectra for 557 objects with $b_J\le
19.45$ and $S\ge 1$~mJy, corresponding to 53 per cent of the original sample. 
This apparent incompleteness is 
merely due to incomplete sky coverage of the spectroscopic survey. 
Neither radio nor magnitude biases have been found in the determination of 
the optical and spectroscopic counterparts of FIRST radio sources. 

\begin{figure*}[t]
\vspace{8cm}  
\includegraphics{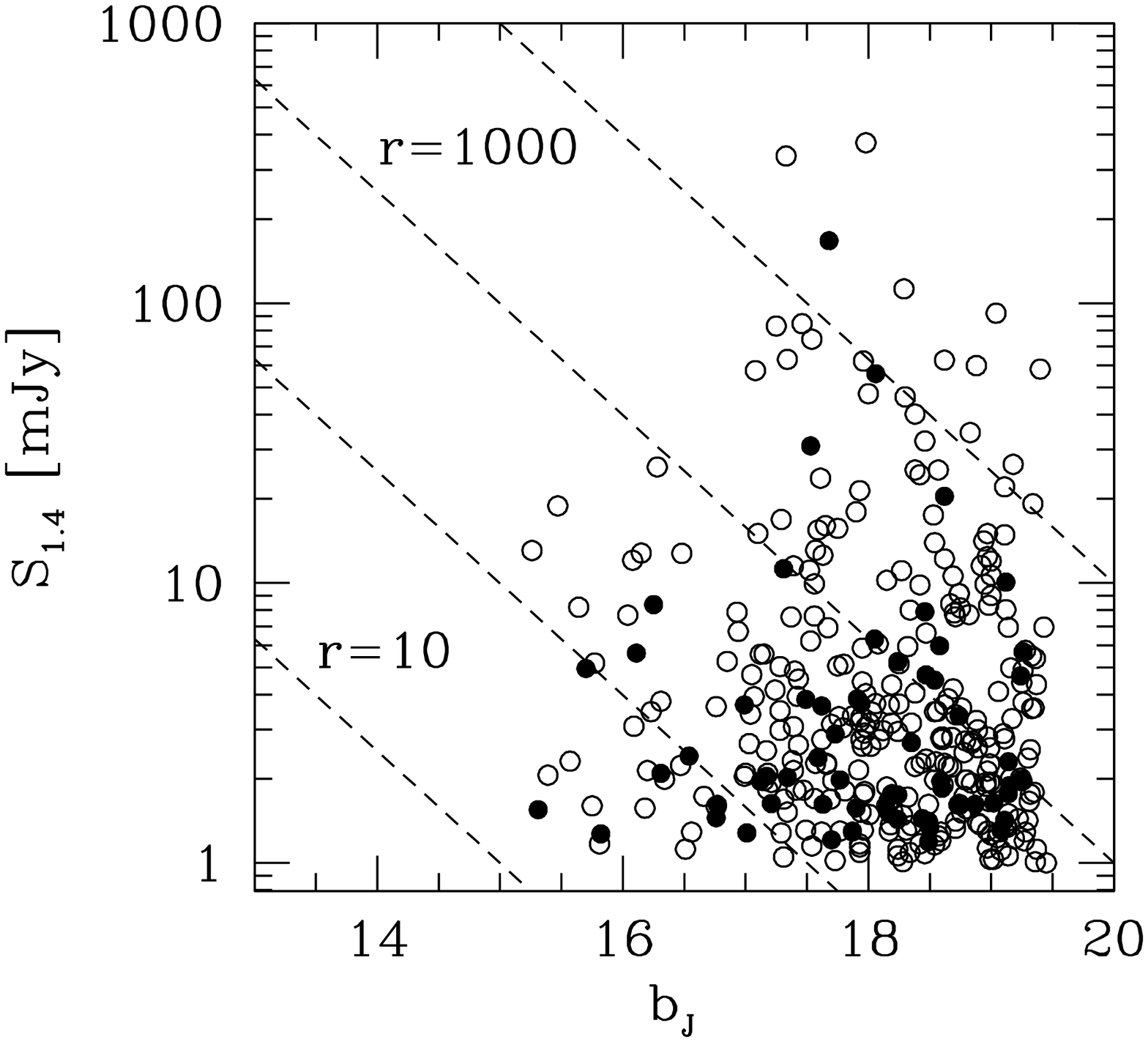}
\includegraphics{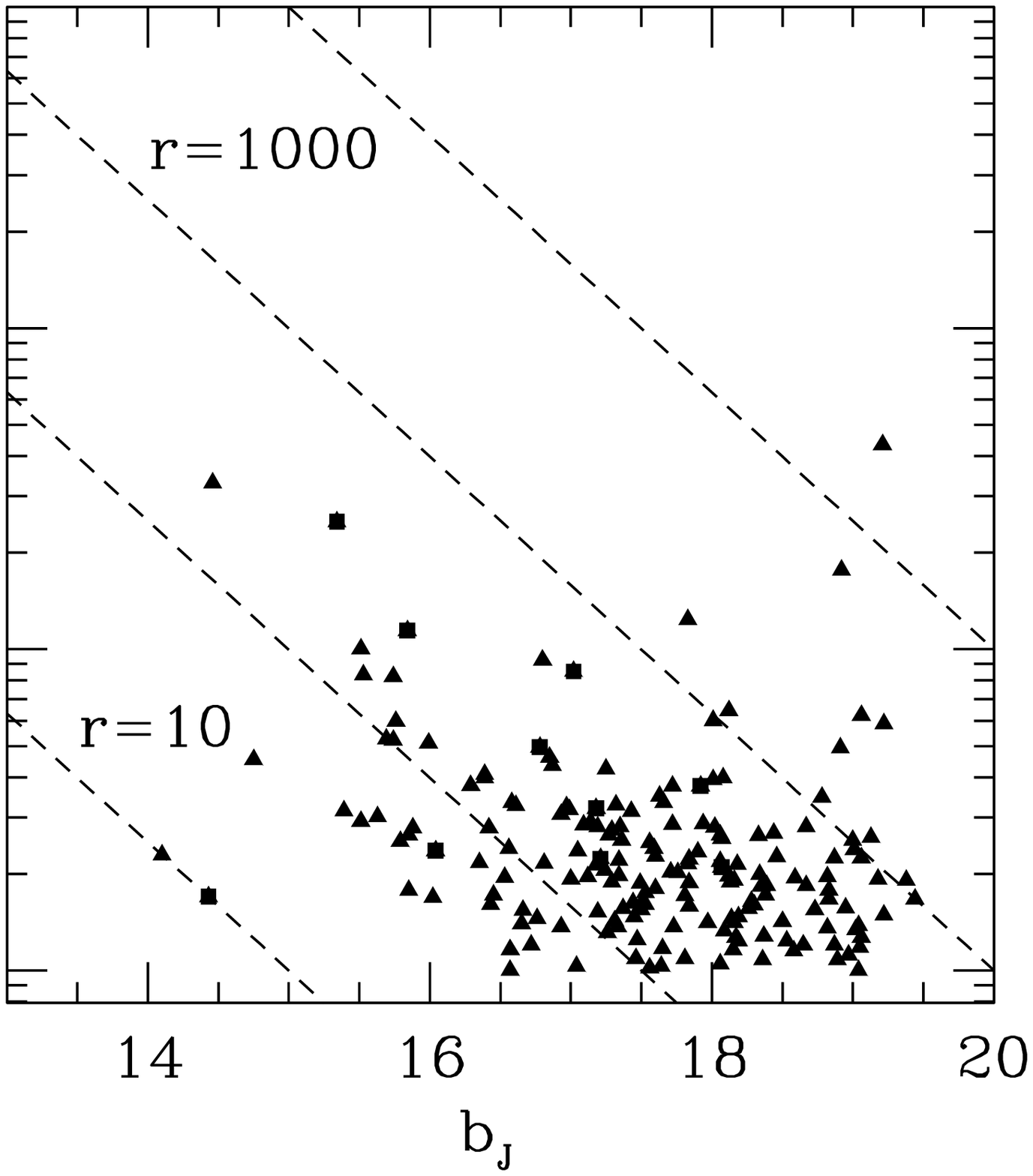}
\includegraphics{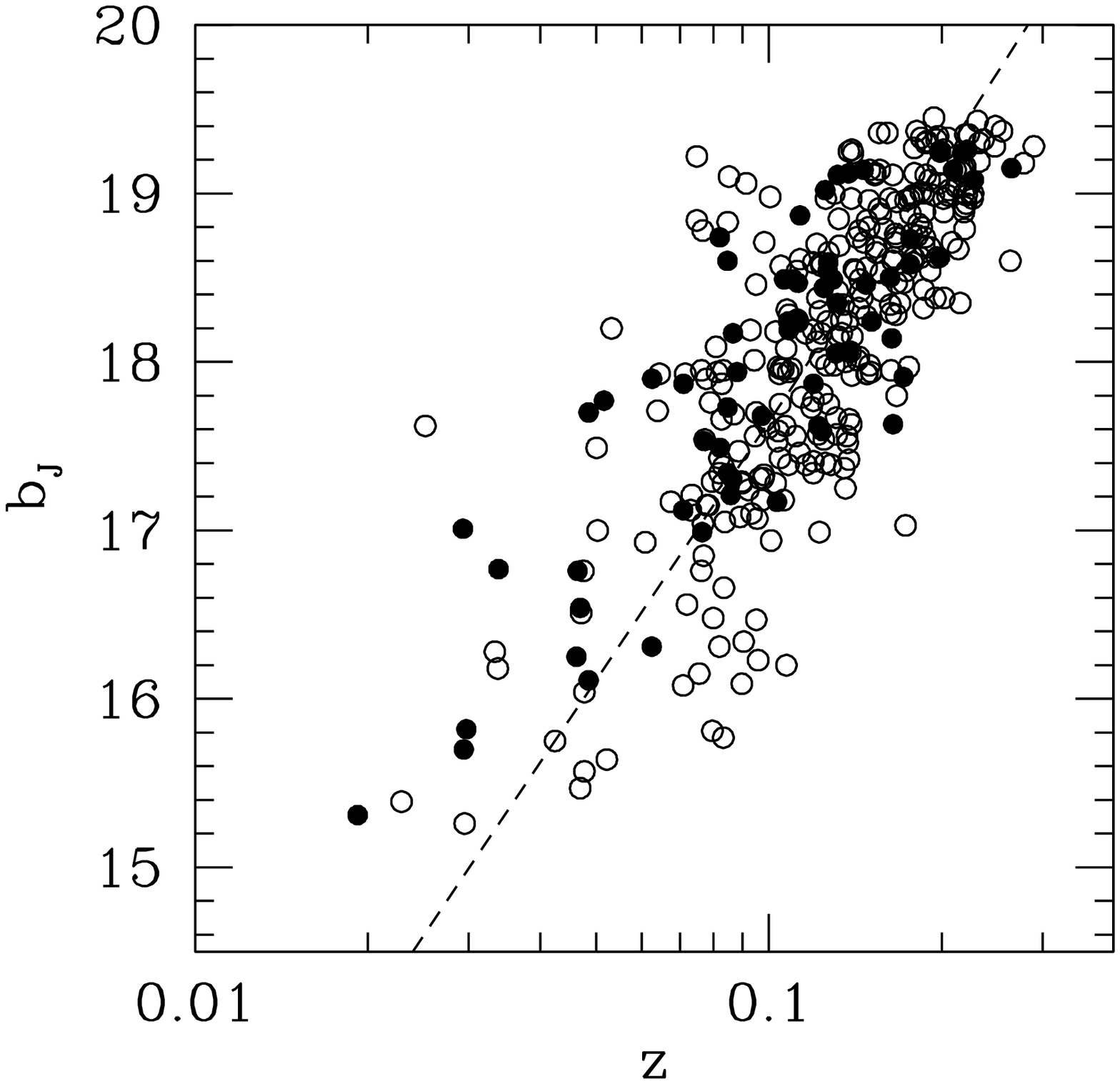}
\includegraphics{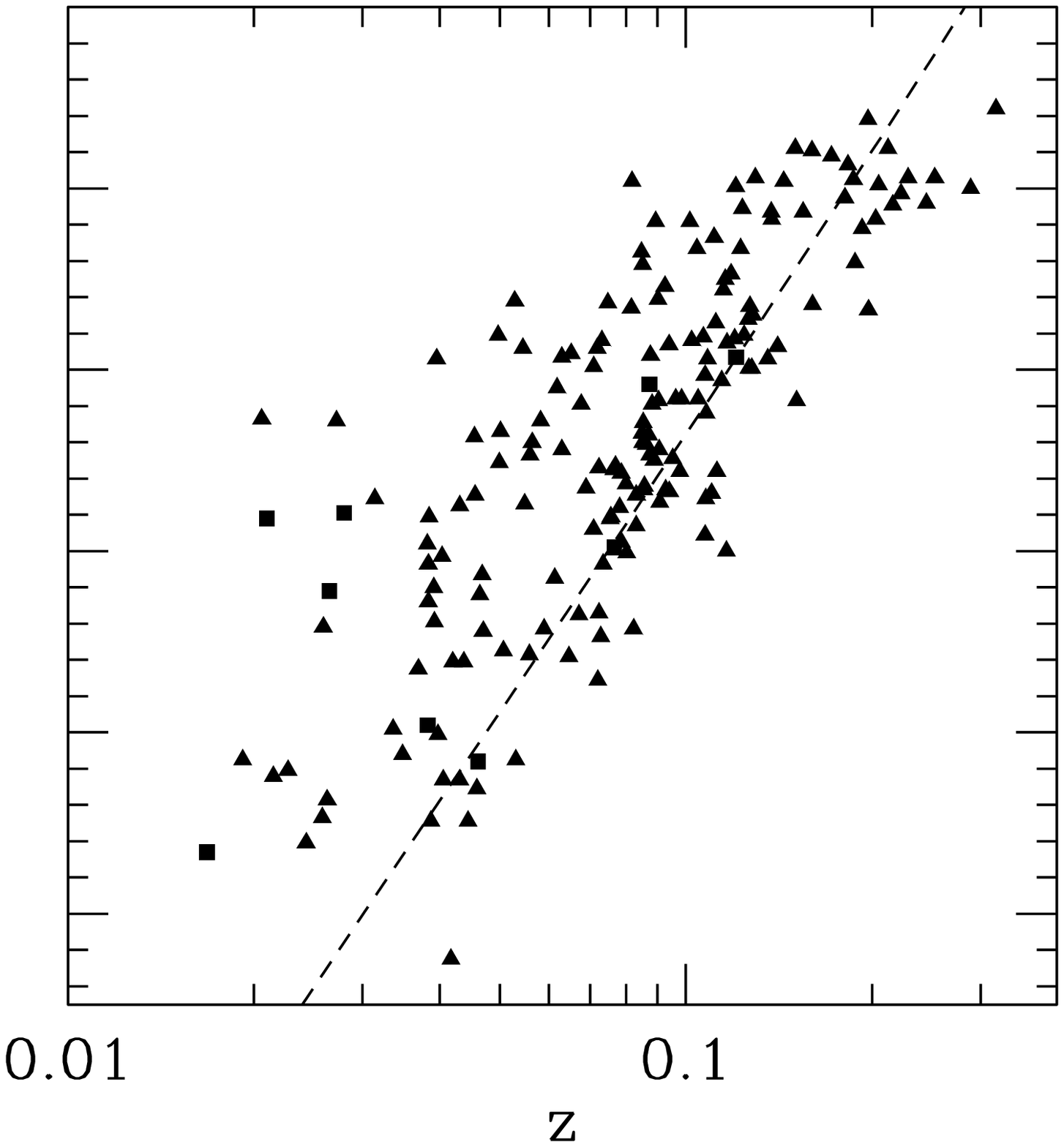}
\caption{$b_J$ magnitudes versus radio flux $S$ at 1.4 GHz (top panels) and 
redshift (bottom panels). Plots are derived for early-type 
(empty circles) and E+AGN (filled circles) galaxies, 
late-type (filled triangles) galaxies and starbursts (filled squares).
The dashed lines in the top panels correspond to constant values of the 
radio-to-optical ratios $r=1-10^4$, while those in the $b_J-z$ 
plots represent the best fit to the data obtained for 
the population of early-type galaxies (see text for details).
\label{fig:F_B}}   
\end{figure*} 

\section{Properties of the Sample}
Classes for the optical counterparts of radio sources
 have been assigned on the basis of their 2dF spectra which
 allowed galaxies to be divided into 4 broad categories.
As Fig.1 shows, different classes  
occupy different regions on the $S-b_J$ and $b_J$-z planes, reflecting the  
different intrinsic characteristics of the populations 
which we summarize here:\\
1) {\bf Early-type galaxies} show spectra dominated by continuum which is much 
stronger than the intensity of any emission line.
This class comprises 289 objects and makes up 52 per cent of the whole
spectroscopic sample. Sources belonging to this population tend to have 
relatively high values for the radio-to-optical ratio, 
$r=S\times 10^{(b_j-12.5)/2.5}$
($100\la r\la 10^4$) and are preferentially found for redshifts
$z \ga 0.1$. The majority of these sources show rather red colours
($b_J-R\ga 1$), with a tendency to be redder at larger look-back
times. Their radio fluxes lie in the range $1\la S/{\rm mJy}
\la 10$, even though objects are found up to $S\sim 400$~mJy, and
optically they appear as relatively faint (about 60 per cent of the
sources has $b_J\ge 18$).  The $b_J-z$ relation in this case
(represented by the dashed line in the bottom-left panel of Fig. 1)
is well described by the expression $b_J-M_B=-5+ 5\;{\rm log_{10}}d_L({\rm pc})
$,  ($d_L$ is the luminosity distance), with $M_B\simeq -21.3$ 
for an $h_0=0.5$, $\Omega_0=1$ universe, 
showing that passive radio galaxies are reliable standard candles.\\
2) {\bf E+AGN-type galaxies} show spectra typical of early-types 
plus the presence of (narrow) emission lines which 
indicate the presence of an active nucleus.
There are 61 objects in this class, corresponding to 11 per cent of
the spectroscopic sample.  These sources are directly connected to the
class of early-type galaxies, even though they show characteristics
that are intermediate between pure AGN-fuelled sources and
star-forming galaxies. Optically they appear as rather faint and
closely follow the standard-candle relationship found for early-type
galaxies.  However their radio-to-optical ratios are in general as low
as those obtained for late-type galaxies.  Their radio fluxes are in
general quite low and their $b_J-R$ colours are
uniformly distributed between about 0 and 2.\\
3)-4) {\bf Late-type galaxies} and {\bf Starbursts} have spectra which show 
strong emission lines characteristic of star-formation activity, 
either together with 
a detectable (late-type) or missing (starbursts) continuum.
This class comprises 177 objects (including 10 starbursts), which is $\sim 30$
per cent of the spectroscopic sample.  The radio-to-optical ratios $r$
for these sources have values in general
smaller than those found for early-type galaxies. Their radio fluxes
are rarely brighter than $S\simeq 5$~mJy and optically they have
bright apparent magnitudes (very few objects are found with $b_J\ga
18.5$).  These sources have quite blue colours, $-2\la b_J-R \la
1$, and are mostly local -- the majority of them being located within
$z \simeq 0.1$. Furthermore, they show a weaker correlation between
$b_J$ magnitudes and redshift than found for the previous classes 
of objects, and do not 
follow the $b_J-z$ relation
found for early-type galaxies (illustrated by the dashed line in the
 bottom panels of Fig. 1).\\
Note that we also found 18 Seyfert 2 and 4 Seyfert 1 galaxies. 
For further discussion and the whole list of objects see Magliocchetti 
et al. (2001).

Some information on the above classes of sources can also be derived from the 
few optical images showing resolved structures. For instance, it is 
interesting to note that the majority of the interacting/merging systems 
seems to be
associated with early-type spectra, typical of pure AGN-fuelled sources, 
suggesting that merging, under appropriate conditions, can trigger AGN 
activity even at low redshifts. Also, as expected, irregulars and spirals 
preferentially show spectra typical of late-type galaxies; signatures 
of interaction and/or merging are seen for members of this latter population 
as well as for E+AGN galaxies. 
Finally, radio images show that there is a clear trend for
extended/sub-structured sources to be associated with absorption
systems (i.e. early-type galaxies).

\begin{figure}[t]
\vspace{5.5cm}  
\includegraphics{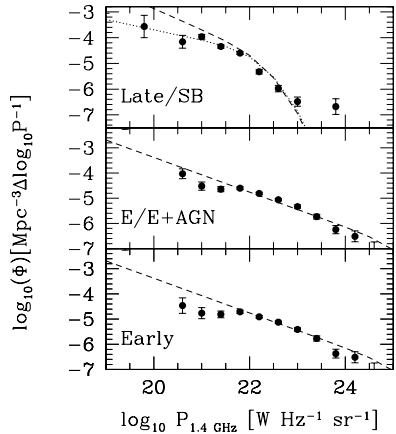}
\caption{Local radio luminosity function at 1.4 GHz for late-type+starburst 
galaxies 
(top panel), early and E+AGN galaxies (middle panel) and early-type 
galaxies only (bottom panel).
\label{fig:lum_fcntype}}   
\end{figure}

Radio powers have subsequently been calculated according to the 
relation $P_{1.4\rm GHz}=
S_{1.4\rm GHz} D^2 (1+z)^{3+\alpha}$, with $D$ angular
diameter distance and $\alpha$  spectral index of the radio
emission, and the local radio LF for objects in the spectroscopic sample 
has been derived by grouping the sources in bins 
of width $\Delta {\rm log_{10}} P=0.4$, according to the expression
$\Phi(P)=\sum_i N_i(P, P+\Delta P)/V_{\rm max}^i(P)$
where $N_i$ is the number of objects with luminosities between $P$ and
$P+\Delta P$ and $V_{\rm max}^i(P)$ is the maximum comoving volume within
which an object could have been detected above the radio-flux and
magnitude limit of the survey. We set $S=1$~mJy for the 
radio-flux limit, while $b_J=19.45$ is the chosen magnitude limit and the 
results for the local LF have been corrected for incompleteness effects 
coming from both the radio and the spectroscopic surveys. \\
This was done for each population taken individually.
Fig.2 shows the results for the early-type
(lower panel), early plus E+AGN (middle panel) and late-type+starburst (top
panel) galaxies. The dashed line in the bottom and middle panels of
the Figure is the predicted LF for steep spectrum FR~I+FR~II sources as 
given by Dunlop \& Peacock (1990) 
under the assumption of pure luminosity evolution. 
In this case the agreement with the data
is very good down to powers $P\simeq 10^{20.5} \whzsr$, 
especially if one includes all the objects which show
spectra typical of early-type galaxies, regardless the presence of
emission lines of AGN origin. 
The LF for late-type galaxies and starbursts is shown in the top panel 
of Fig.2, and features a break at about $P\simeq 10^{22} \whzsr$, beyond 
which the contribution of this class of objects becomes rapidly negligible. 
When it comes to a comparison between observed and predicted LF for
this latter case, one has that 
a good description of the data is provided by the Rowan-Robinson et al. (1993) 
model (dashed line in the top panel of Fig.2), which can 
correctly reproduce both the broken power-law behaviour and the break 
luminosity, therefore supporting the assumption of these authors for late-type 
radio galaxies to be identified with the population of dusty spirals 
and starbursts observed at 60~$\mu$m. 
However, the Rowan-Robinson et al. (1993) predictions result in an 
over-estimate of the number of faint (i.e. $P\la 10^{21} \whzsr$) radio 
sources, while the best fit is provided by 
assuming a shallower faint-end slope for the local LF, as illustrated by 
the dotted line in Fig.2 (see Magliocchetti et 
al. 2001 for more details).

\end{document}